\documentclass[10pt]{article}
\usepackage{amsmath}
\usepackage{amsfonts}
\usepackage{amssymb}
\usepackage{amscd}
\newcommand{\Adg}{\mathrm{Ad}_g \,}

\newcommand{\shone}{\left[1\right]}
\newcommand*\samethanks[1][\value{footnote}]{\footnotemark[#1]}
\begin{document}
\author{Anton Alekseev \thanks{Section de math\'ematiques,
Universit\'e de Gen\`eve,
2-4 rue du Li\`evre,
c.p. 64, 1211 Gen\`eve 4,
Switzerland} \and
Yves Barmaz \samethanks \and
Pavel Mnev \thanks{Institut f\"ur Mathematik,
Universit\"at Z\"urich,
Winterthurerstrasse 190,
CH-8057, Z\"urich, Switzerland}}
\title{Chern-Simons Theory with Wilson Lines and Boundary in the BV-BFV Formalism}
\maketitle
\begin{abstract}
We consider the Chern-Simons theory with Wilson lines  in 3D and in 1D in the BV-BFV formalism of Cattaneo-Mnev-Reshetikhin. In particular, we allow
for Wilson lines to end on the boundary of the space-time manifold.
In the toy model of 1D Chern-Simons theory, the quantized BFV boundary action
coincides with the Kostant cubic Dirac operator which plays an important
role in representation theory. In the case of 3D Chern-Simons theory, the
boundary action turns out to be the odd (degree 1) version of the $BF$ model with source terms for the $B$ field at the points where the Wilson lines meet the boundary. The boundary
space of states arising as the cohomology of the quantized BFV action coincides
with the space of conformal blocks of the corresponding WZW model.
\end{abstract}
\section{Introduction}
%\textbf{In the framework of the Batalin-Vilkovisky formalism for the quantization of field theories with degeneracies, the situation where the source manifold has a boundary is not yet well understood.} So far, one has mostly relied on a careful choice of boundary conditions to satisfy the master equation. Nevertheless,
In the case of complicated space-time topology, a promising approach to quantization of general field theories involves cutting the space-time manifold into simple pieces, where the problem is more easily solved, and then gluing back the individual elements to obtain the final answer. This method was proposed by Atiyah and successfully applied by Witten in \cite{WittenCS} to study the quantization of the Chern-Simons theory with Wilson lines (cf. also \cite{Froehlich-King}). Following this idea, a systematic program to understand quantization in the Batalin-Vilkovisky formalism for field theories with degeneracies on manifolds with boundaries has been initiated in \cite{CattaneoMnevReshetikhin}. As a part of the construction, a Batalin-Fradkin-Vilkovisky model is associated to the boundary of the space-time manifold. The canonical quantization of this boundary BFV model provides a space of boundary states, together with a cohomological invariance condition that defines the admissible quantum states of the theory among all boundary states. In the case of quantum field theories on manifolds without boundary, the partition function and other correlation functions are complex-valued. In the presence of a boundary, the correlators of the bulk theory take values in this boundary space of states.

The aim of this paper is to apply the
BV-BFV formalism of \cite{CattaneoMnevReshetikhin} to the Chern-Simons theory on manifolds with boundary, with Wilson lines ending on the boundary. The BV formulation of this theory on closed manifolds is well understood (and served as the motivating example for the AKSZ construction). We will also consider the one-dimensional Chern-Simons model, obtained when the AKSZ construction is carried out in one dimension.

We include in our construction Wilson lines which may end on the boundary of the manifold. This requires some extra work in BV-BFV formalism. Our treatment is based on the path integral representation for Wilson loops suggested in \cite{QuantSymplOrbits}, \cite{Diakonov-Petrov}. This is also an example of a more general construction of observables for AKSZ sigma models proposed in \cite{AKSZObs}.

We compare our answers with those obtained using the geometric quantization for the boundary \cite{WittenCS} and using canonical quantization \cite{CSgenus0}, \cite{CSgenus1}.

One of our main results is the boundary BFV action for Chern-Simons theory with Wilson lines, which has the form of an odd (degree 1) version of $BF$ action modified by source terms for the $B$ field at points where the Wilson lines meet the boundary. We also consider the toy model of one-dimensional Chern-Simons theory and derive the corresponding boundary action. Its quantization coincides with Kostant cubic Dirac operator. We compare the BV-BFV results for the one-dimensional model with the ones obtained in  \cite{1DCSAlekseevMnev} on segments, and also see how Wilson lines can be added to the one-dimensional model. In the three-dimensional case, the boundary space of states, arising as the cohomology of the quantized BFV action, coincides with the space of conformal blocks of the WZW model on the boundary (in the picture of \cite{CSgenus0}).

%Since the answer of the quantization problem of the Chern-Simons theory is known within the geometric quantization picture \cite{WittenCS} and the canonical quantization picture %\cite{CSgenus0}, \cite{CSgenus1}, we will get a direct interpretation of the one we obtain in the BV-BFV picture. However, these previous results all include Wilson lines in their %treatment, so we will need to take into account such observables as well. It turns out that they can be cast into an extension of the AKSZ construction: the space of fields needs to %be augmented with fields living on the curve supporting the Wilson line, associated to some coadjoint orbit of the gauge group, and the path-ordered holonomy (i.e. the Wilson line) %can be replaced by an additional term in the action. We will be able to compare the BV-BFV results for the one-dimensional model with the ones obtained in  \cite{1DCSAlekseevMnev} on %segments, and also see how Wilson lines can be added to the one-dimensional model.

We begin in section \ref{WilsonLinesBV} with the treatment of Wilson lines in the BV formalism.
%We keep the discussion general, i.e. we consider any gauge theory and do not restrict to the simplest case of Chern-Simons theory, as these results might be useful for readers interested in other models where Wilson lines are relevant.
We also provide a short introduction to the relevant aspects of the BV formalism. In section \ref{3DCSWL} we describe the BV formulation of the Chern-Simons theory with Wilson lines, applying the AKSZ construction to this special setting. Then we proceed to explain how the bulk model gets supplemented with a boundary BFV theory if the underlying manifold has a boundary. We repeat this procedure in section \ref{1DCSWL} for the one-dimensional Chern-Simons model. The $\mathbb{Z}_2$-grading that replaces the usual $\mathbb{Z}$-grading in this case leads to certain subtleties with the master equation. In section \ref{BoundaryQuantumStates} we present the quantization of the boundary BFV models and describe the arising spaces of quantum states, that we compare with known results for the quantization of the involved models.

\subsection*{Acknowledgements}
Research of A.A. and Y.B. was supported in part by the grants number 140985 and 141329 of the Swiss National Science Foundation. P.M. acknowledges partial support by RFBR grant 11-01-00570-a and by SNF grant 200021\_137595.

\section{Wilson lines in the BV formalism} \label{WilsonLinesBV}
In this section, we start with a brief introduction to the BV formalism. The main point is to incorporate the Wilson line observables in this approach.
\subsection{A short introduction to the BV formalism}
We know that the path integral in quantum field theories is not well defined if the classical action $S_{cl}$ defined over the space of classical fields $\mathcal{F}_{cl}$ is degenerate, for instance due to gauge symmetries. The Batalin-Vilkovisky formalism provides a general method for the perturbative calculation of partition functions and correlators.

In the BV formalism, the space of fields is augmented to a BV space of fields $\mathcal{F}_{BV}$, a graded infinite-dimensional manifold equipped with a symplectic structure  $\Omega_{BV}$ of degree -1 called the BV structure. The grading (usually $\mathbb{Z}$, sometimes $\mathbb{Z}_2$) is commonly referred to as ``ghost number'', in relation with the Faddeev-Popov prescription. The BV bracket is defined as the Poisson bracket obtained by inverting the BV structure,
\[
\left\lbrace F,G \right\rbrace = \Omega_{BV}^{-1}(\delta F,\delta G),
\]
and obviously has a ghost number 1. Note that the variational operator $\delta$ can be interpreted as a de Rham differential in the space of fields. In many cases of interest, the BV space of fields is a cotangent bundle where the degree of the fibers is shifted to $-1$, which ensures its canonical symplectic form has the proper degree. Coordinates along the cotangent fibers are then called antifields. In the case of gauge theories, where the degeneracy arises under the action of a gauge group, the BV space of fields is simply the shifted cotangent bundle of the BRST space of fields which contains all classical fields as well as the ghosts parametrizing the gauge symmetries (basically the infinitesimal gauge parameters with a ghost number shifted by one), $\mathcal{F}_{BV}=T^\ast\left[-1\right]\mathcal{F}_{BRST}$.

At the classical level, infinitesimal gauge transformations and the classical action can be used to construct a differential acting on the functionals on the BV space of fields. Geometrically, this differential corresponds to a cohomological vector field $Q$ on $\mathcal{F}_{BV}$. Moreover, $Q$ is a Hamiltonian vector field,
\[
\imath_Q \Omega_{BV} = \delta S_{BV},
\]
 with the Hamiltonian function being the BV action $S_{BV}$ that reduces to the classical action when all antifields are set to zero. The condition $Q^2=0$ follows from the classical master equation $\left\lbrace S_{BV},S_{BV}\right\rbrace =0$, and determining the BV formulation of a given theory amounts to determining an extension of the classical action to the BV space of fields that satisfies this classical master equation.
\subsection{The AKSZ construction}
While it is usually difficult to find the BV formulation of a given field theory with a degenerate action, the study of the geometric interpretation of the classical master equation in \cite{AKSZ} led to an insightful procedure to construct solutions thereof, called the AKSZ construction after its authors. A formalized more recent treatment can be found in \cite{AKSZCattaneoFelder}.

In this construction, the target space of the theory is a graded manifold $Y$ equipped with a symplectic structure $\omega_Y$ of degree $n-1$ and a compatible cohomological vector field $Q_Y$ of degree 1, in the sense that it preserves the symplectic structure, $\mathcal{L}_{Q_Y}\omega_Y=0$. We also want $\omega_Y$ to be associated to a Liouville one-form $\alpha_Y$ (of degree $n-1$ as well), namely $\omega_Y=\delta\alpha_Y$. Here $\delta$ denotes the de Rham exterior derivative on $Y$, while we keep the usual $d$ for the one on the source manifold $N$ of the model. For $n\neq 0$ (see for instance \cite{Roytenberg} for details), $Q_Y$ can be shown to be Hamiltonian, i.e. there exists a function $\Theta_Y$ of degree $n$ on $Y$ such that $\imath_{Q_Y}\omega_Y=\delta\Theta_Y$. Like in the BV formalism, the nilpotency of $Q_Y$ follows from the condition $\left\lbrace \Theta_Y,\Theta_Y\right\rbrace_Y =0$, where the curly braces with a subscript $Y$ denote the Poisson bracket on $Y$ associated to its symplectic structure.

The BV space of fields is then given by maps between the odd tangent bundle of some $n$-dimensional manifold $N$ and the graded manifold $Y$,
\[
\mathcal{F}_{AKSZ}=\mathrm{Map}( T\left[1\right]N ,Y).
\]
The odd tangent bundle is naturally equipped with a cohomological vector field, the de Rham vector field, which can be expressed as
\[
D=\theta^\mu \frac{\partial}{\partial x^\mu}
\]
in coordinates $x^\mu$ of the base manifold $N$ and $\theta^\mu$ of the odd fibers. Notice also that real-valued functions on $T\left[1\right]N$ can be interpreted as differential forms on $N$ (by expanding the function in powers of $\theta^\mu$),
\[
C^\infty(T\left[1\right]N,\mathbb{R}) \simeq \Omega^\bullet(N),
\]
which allows to define a canonical measure $\mu$ on $T\left[1\right]N$: the Berezinian integration along all odd fibers simply extracts the top-form out of this expansion and it remains to integrate it over the base $N$.

Roughly, the idea behind the AKSZ construction involves lifting the symplectic structure $\omega_Y$ from the target space to define the BV structure $\Omega_{\mathrm{AKSZ}}$ on the space of fields. In effect, we replace functions and differential forms on $Y$ by functionals on the space of fields with values in differential forms on $N$ and their variations (which also explains the choice of $\delta$ to denote the exterior derivative on $Y$), and we integrate over the source space $T\left[1\right]N$ using its canonical measure $\mu$,
\begin{equation} \label{AKSZBVstr}
\Omega_{\mathrm{AKSZ}} = \int_{T\left[1\right]N} \mu\, \tilde{\omega}_Y.
\end{equation}
The tilde denotes the extension from function on $Y$ to functional on the space of fields. The Berezinian integration along the fibers will lower the ghost number of $\omega_Y$ from $n-1$ to $-1$ as required.

In a second stage we need to lift the cohomological vector field $Q_Y$ on the target-space $Y$ as well as the de Rham vector field $D$ on the source-space $T\shone N$ to the space of fields, and combine them to form the BV cohomological vector field $Q$ discussed above that will happen to be Hamiltonian. Its generating functional is nothing but the BV-AKSZ action
\begin{equation} \label{AKSZaction}
S_{\mathrm{AKSZ}} = \int_{T\left[1\right]N} \mu\ \left( \imath_{Q_D}\tilde{\alpha}_Y + \tilde{\Theta}_Y \right),
\end{equation}
where $Q_D = \Sigma_i D\phi^i \frac{\delta}{\delta \phi^i}$ is the lift of the de Rham vector field ($\phi^i$ denotes generic coordinates on the space of fields). This AKSZ action automatically solves the classical master equation as a consequence of the integrability condition on $\Theta_Y$ and the fact that the integral of exact forms vanishes provided $\partial N=\emptyset$.

As an example of the AKSZ construction, we derive here the BV formulation of the Chern-Simons theory, which corresponds to the special case $n=3$ with $Y=\mathfrak{g}\shone$, where $\mathfrak{g}$ is a Lie algebra equipped with an invariant scalar product. As required, the target space $\mathfrak{g}\left[1\right]$ supports a symplectic structure of degree 2 and a Hamiltonian cohomological vector field of degree 1 (sometimes called a Q-structure). If we denote with $\delta$ the exterior derivative on $\mathfrak{g}\left[1\right]$ and $\psi$ a generic element, the symplectic form, its Liouville potential and the Hamiltonian of the cohomological vector field respectiviely can be written as
\begin{eqnarray}
\omega_{\mathfrak{g}\shone} &=& -\frac{1}{2} \left( \delta\psi,\delta\psi \right), \\
\alpha_{\mathfrak{g}\shone} &=& -\frac{1}{2} \left( \psi,\delta\psi \right), \\
\Theta_{\mathfrak{g}\shone} &=& -\frac{1}{6} \left( \psi,\left[\psi,\psi\right] \right).
\end{eqnarray}
Note that Grassmanian variables $\psi$ anticommute, but so do differential forms of odd degree, which explains why $\omega_{\mathfrak{g}\shone}$ may be built out of a symmetric product.

If we use coordinates $x^\mu,\mu=1,2,3$ on $N$ and corresponding Grassmanian coordinates $\theta^\mu$ on the odd fibres of $T\shone N$, we can decompose the fields $\mathbf{A}\in\mathrm{Map}( T\shone N ,\mathfrak{g}\shone)$ into $\mathfrak{g}$-valued differential forms of various degrees and grading,
\begin{equation} \label{decompo3D}
\mathbf{A}=\gamma + A_\mu\theta^\mu + \frac{1}{2} A^+_{\mu\nu}\theta^\mu\theta^\nu + \frac{1}{6} \gamma^+_{\mu\nu\sigma}\theta^\mu\theta^\nu\theta^\sigma,
\end{equation}
specifically
\begin{displaymath}
\begin{array}{lcl}
\gamma & \in & \mathrm{Map}(N,\mathfrak{g}[1]), \\
A & \in & \Gamma( T^\ast N \otimes \mathfrak{g}), \\
A^+ & \in & \Gamma( \bigwedge^2 T^\ast N \otimes \mathfrak{g}[-1]), \\
\gamma^+ & \in & \Gamma( \bigwedge^3 T^\ast N \otimes \mathfrak{g}[-2]).
\end{array}
\end{displaymath}
These fields are endowed with two gradings, namely the ghost-grading (that stands in square brackets when non-zero) and the degree as a differential form. Their sum, the total degree, should amount to 1, since each $\theta^\mu$ has a ghost number 1, and all terms in the decomposition (\ref{decompo3D}) should have the same total ghost number of 1. As usual, the fields of ghost number 0 are the classical fields, here a $\mathfrak{g}$-valued connection $A$, and the fields of ghost number 1 are simply called ghosts. The other two fields are their antifields (which in the BV formalism means canonically conjugated), as is clear when one computes the BV structure,
\begin{equation} \label{3D_BV_form}
\begin{split}
\Omega_{\mathrm{BV}}^{\mathrm{CS}} &= \int_{T\shone N}\mu \ \tilde{\omega}_{\mathfrak{g}\shone}= - \int_{T\shone N}\mu \left( \delta\mathbf{A},\delta\mathbf{A} \right) \\
&= \int_N \left( \left( \delta \gamma^+,\delta \gamma \right) - \left( \delta A^+,\delta A \right) \right) = \int_N \left( -\left( \delta \gamma,\delta \gamma^+ \right) + \left( \delta A,\delta A^+ \right) \right).
\end{split}
\end{equation}

Note that the commutation rules for the fields (which are simultaneously functions on $\mathcal{F}$ and differential forms on $N$) are determined by the total degree (the de Rham degree of the differential form plus the ghost number): two fields of odd total degree anti-commute, and commute if at least one field has even total degree.

It remains to compute the BV action, which is straightforward in the AKSZ scheme,
\begin{equation}   \label{Scsbv}
\begin{split}
S_{\mathrm{BV}}^{\mathrm{CS}} &= \int_{T\shone N} \mu \left( \imath_{Q_D} \tilde{\alpha}_{\mathfrak{g}\shone} + \tilde{\Theta}_{\mathfrak{g}\shone} \right) \\
&= \int_{T\shone N} \mu \left( \left( \mathbf{A},D\mathbf{A} \right) + \frac{1}{6} \left( \mathbf{A},\left[\mathbf{A},\mathbf{A}\right] \right) \right) \\
&= \int_N \left( \frac{1}{2} \left( A, dA \right) + \frac{1}{6} \left( A,\left[ A,A\right] \right) - \left( A^+ , d\gamma + \left[ A,\gamma\right] \right) + \left( \gamma^+,\frac{1}{2}\left[ \gamma,\gamma \right] \right) \right).
\end{split}
\end{equation}
In the two terms involving only physical fields, we recognize the classical action of the Chern-Simons theory. The other terms complete the BV action, and by construction, it is clear that it satisfies the classical master equation. Nevertheless, we will show it explicitly, mainly to present an example of calculations in the space of fields, on which we will rely in the rest of this paper.
\subsection{Calculations in the BV formalism}
%(Should we rather put it as an annex?)

First of all, we need to find the BV bracket, simply by inverting the BV structure (\ref{3D_BV_form}), without forgetting that the product between two differential forms in the space of fields really means the exterior product,
\begin{displaymath}
\delta \phi^+ \delta \phi = \delta \phi^+ \wedge \delta \phi = \delta \phi^+ \otimes \delta \phi \pm \delta \phi \otimes \delta \phi^+,
\end{displaymath}
where the sign depends on the commutation rules between $\phi$ and $\phi^+$. We find the following expression for the BV bracket of two functionals $F_1$ and $F_2$,
\begin{equation}
\begin{split}
\left\lbrace F_1,F_2\right\rbrace = & \int_N \left(
\left( \frac{F_1 \overleftarrow{\delta}}{\delta \gamma}, \frac{\overrightarrow{\delta}F_2}{\delta \gamma^+}\right)
- \left( \frac{F_1 \overleftarrow{\delta}}{\delta \gamma^+}, \frac{\overrightarrow{\delta}F_2}{\delta \gamma}\right) \right. \\
& \qquad \left.
- \left( \frac{F_1 \overleftarrow{\delta}}{\delta A}, \frac{\overrightarrow{\delta}F_2}{\delta A^+}\right)
+ \left( \frac{F_1 \overleftarrow{\delta}}{\delta A^+}, \frac{\overrightarrow{\delta}F_2}{\delta A}\right) \right)
\end{split}
\end{equation}
where the functional derivatives $\frac{\overrightarrow{\delta}}{\delta\phi}$ and $\frac{\overleftarrow{\delta}}{\delta\phi}$ for $\phi\in\left\lbrace A, A^+, \gamma, \gamma^+, g^+\right\rbrace$ are the duals of the differentials $\delta\phi$ in the space of fields (which can be interpreted as variations of fields in the framework of variational calculus). We need to make the difference between right- and left-derivatives due to the commutation rules that depend on ghost numbers and degrees of differential forms on $N$.

All the fields of the Chern-Simons model are $\mathfrak{g}$-valued, so taking the functional derivative of a real-valued functional $F$ on $\mathcal{F}$ by one of these fields should produce a $\mathfrak{g}^\ast$-valued result, but we can use the non-degenerate scalar product $\left(\cdot,\cdot\right)$ to identify $\mathfrak{g}$ with its dual. If $F$ is constructed as an integral, like an action, the left- and right-derivatives by a field $\phi\in\left\lbrace A, A^+, \gamma, \gamma^+, g^+\right\rbrace$ can be defined as the components of the exterior derivative with respect to the local frame induced by these coordinate-fields of $\mathcal{F}$,
\begin{displaymath}
\delta F(\phi_1,\dots,\phi_n) = \int_{\partial N} \sum_{j=1}^n \left(\delta\phi_j,\frac{\overrightarrow{\delta}F}{\delta \phi_j}\right) = \int_{\partial N} \sum_{j=1}^n \left(\frac{F\overleftarrow{\delta}}{\delta \phi_j},\delta\phi_j \right).
\end{displaymath}
If on the other hand $F$ is a local functional, we may still express it as an integral provided we filter its position with a Dirac distribution, a distribution that will stick to the functional derivative.

At a later stage, we will need to consider Lie group-valued fields of the form $g\in\mathrm{Map}(N,G)$. The problem with such a field is that its variation does not take value in $\mathfrak{g}$, but rather in the tangent space at $g$ of the Lie group $G$. Natural coupling with the other $\mathfrak{g}$-valued fields via the invariant scalar product involves the right multiplication by $g^{-1}$ to bring it back to the Lie algebra, explicitly $\delta g\, g^{-1}$. The dual derivative $\frac{\delta}{\delta g}$ assumes its value in $T^\ast_{g^{-1}} G$, which is isomorphic to $T_{g^{-1}} G$ thanks to the invariant non-degenerate scalar product, and we need to apply this time left-multiplication by $g$ to get back to the Lie algebra. If the functional $F$ depends also on $g$, we find for the derivative
\begin{displaymath}
\delta F(\phi_1,\dots,\phi_n, g) = \int_{\partial N} \sum_{j=1}^n \left(\delta\phi_j,\frac{\overrightarrow{\delta}F}{\delta \phi_j}\right) +\left( \delta g\, g^{-1}, g \frac{\overrightarrow{\delta}F}{\delta g}\right).
\end{displaymath}

To compute $\left\lbrace S_{\mathrm{BV}}^{\mathrm{CS}},S_{\mathrm{BV}}^{\mathrm{CS}} \right\rbrace$, we need the derivatives of the Chern-Simons BV action. We find
\begin{equation}
\begin{split}
\delta S_{\mathrm{BV}}^{\mathrm{CS}} = \int_N  & \left(   \left( \delta A,  dA + \frac{1}{2} \left[A,A\right] + \left[A^+,\gamma\right] \right) \right. \\
& + \left( \delta \gamma, -d_A A^+ - \left[\gamma^+,\gamma\right] \right) \\
& + \left. \left( \delta A^+, -d_A\gamma \right) + \left( \delta\gamma^+, \frac{1}{2} \left[\gamma,\gamma\right]\right) \right),
\end{split}
\end{equation}
where we introduced the covariant derivative $d_A=d + \left[A,\cdot\right]$. Note that we need to integrate by parts to find the contribution of the exterior derivatives, such as
\[
\int_N \left( A, \delta dA\right) = \int_N \left( A, d\delta A\right) = - \int_{\partial N} \left(A,\delta A\right) + \int_N \left(dA,\delta A\right),
\]
where the boundary term vanishes for a closed manifold $N$. When we consider source spaces with boundaries, these terms will no longer vanish, and they will contribute to a one-form in the boundary space of fields. For now we can check the classical master equation,
\begin{displaymath}
\begin{split}
\frac{1}{2}\left\lbrace S_{\mathrm{BV}}^{\mathrm{CS}},S_{\mathrm{BV}}^{\mathrm{CS}}\right\rbrace &=
\int_N \left(
\left( \frac{S_{\mathrm{BV}}^{\mathrm{CS}} \overleftarrow{\delta}}{\delta \gamma}, \frac{\overrightarrow{\delta}S_{\mathrm{BV}}^{\mathrm{CS}} }{\delta \gamma^+}\right)
- \left( \frac{S_{\mathrm{BV}}^{\mathrm{CS}} \overleftarrow{\delta}}{\delta A}, \frac{\overrightarrow{\delta}S_{\mathrm{BV}}^{\mathrm{CS}}}{\delta A^+}\right) \right) \\
&= \int_N \left( \left( d_A A^+ + \left[\gamma^+,\gamma\right], \frac{1}{2} \left[\gamma,\gamma\right]\right) \right. \\
& \qquad \quad \left. - \left( dA + \frac{1}{2} \left[A,A\right] + \left[A^+,\gamma\right] , -d_A\gamma \right)\right) \\
&=0.
\end{split}
\end{displaymath}
In the last step we make repeated use of the invariance of the scalar product, the Jacobi identity for $\mathfrak{g}$, and the Stokes theorem.

We are now fully prepared to describe Wilson lines in the BV formalism.
\subsection{Wilson Lines}
In gauge theories, a degeneracy arises under the local action of a Lie group on the space of fields, the gauge group.  In what follows, we will denote by $G$ the gauge group, and $\mathfrak{g}$ its associated Lie algebra. Their so-called gauge field is a connection $A$ in a principal $G$-bundle over some manifold $N$. The gauge symmetry is parametrized in the BV (and BRST) formalism by a ghost field $\gamma\in\mathrm{Map}(N,\mathfrak{g}\left[1\right])$. The BV variation of these two fields depends only on their behavior under gauge transformations and not on the specific type of the underlying ambient theory. We assume that the dynamics and the gauge structure of this ambient theory is encoded in the BV action $S^{\mathrm{amb}}$ and the corresponding BV structure $\Omega^{\mathrm{amb}}$ (both defined as integrals over $N$), and of course that $S^{\mathrm{amb}}$ solves the ambient classical master equation $\left\lbrace S^{\mathrm{amb}},S^{\mathrm{amb}}\right\rbrace_{\mathrm{amb}}=0$. The part of this ambient BV bracket involving the gauge connection and the ghost field relevant for our further investigation is defined by the BV variation of these fields, namely
\begin{equation} \label{gaugetransfo}
\begin{array}{rcccl}
\left\lbrace S^{\mathrm{amb}}, A \right\rbrace &=& Q\,A &=& d_A\gamma, \\
\left\lbrace S^{\mathrm{amb}}, \gamma \right\rbrace &=& Q\, \gamma &=& \frac{1}{2}\left[\gamma,\gamma\right].
\end{array}
\end{equation}

Natural non-local observables to consider in gauge theories are given by Wilson-loops, traces of the holonomy of the connection $A$ along a curve $\Gamma$ embedded in $N$ in given representations of the Lie algebra,
\begin{displaymath}
W_{\Gamma,R}\left[A\right] = \mathrm{Tr}_R\mathrm{Pexp}\left(\int_{\Gamma}A\right),
\end{displaymath}
where P stands for the path-ordering and $R$ labels the representation of $\mathfrak{g}$.

This cumbersome path-ordering can be removed at the price of integrating over all inequivalent gauge transformations along the loop \cite{QuantSymplOrbits},
\begin{equation}
W_{\Gamma,R}\left[A\right] = \int \mathcal{D}g \, \mathrm{exp}\left(\int_\Gamma\langle T_0, g^{-1}Ag + g^{-1}dg\rangle\right).
\end{equation}
The dual algebra element $T_0\in\mathfrak{g}^\ast$ encodes the representation $R$, along the lines of the orbit method \cite{OrbitMethod} that links unitary irreducible representations of Lie groups and their coadjoint orbits. This expression for the Wilson loop can be absorbed into an extended action by adding the auxiliary term
\begin{equation}  \label{wilsonlineterm}
S_{\mathrm{Wilson}} = \int_\Gamma \langle \mathrm{Ad}^\ast_g (T_0), A + dg\,g^{-1}\rangle
\end{equation}
to the ambient action $S^{\mathrm{amb}}$ of the model under consideration. In this last step we replaced the adjoint action on the second factor of the product by the coadjoint action on the first factor, to  emphasize the role of the coadjoint orbit $\mathcal{O}$ of $T_0$.

Now we would like to find a BV formulation of this contribution, so as to obtain a BV action of the full model with Wilson loops. The partition function of such a model with an action extended to take into account a Wilson line as an auxiliary term actually corresponds to the expectation value of this Wilson line in the pure theory,
\begin{displaymath}
Z_{S^{\mathrm{amb}} + S^{\mathrm{aux}}} = \langle W_{\Gamma,R} \rangle_{S^{\mathrm{amb}}}.
\end{displaymath}

We note that the coadjoint orbit $\mathcal{O}$ supports the Kirillov symplectic structure $\omega_{\mathcal{O}}$, of ghost number 0, and that the curve $\Gamma$ carrying the Wilson line has dimension 1, the first two main ingredients for the AKSZ construction for $n=1$. It is thus tempting to try to apply the prescription proposed in \cite{AKSZObs} to construct observables within the AKSZ formalism. Nonetheless, \cite{AKSZObs} treats exclusively the case of an ambient theory of the AKSZ type, whereas we want to consider gauge theories, with the sole requirement that their space of fields contains a gauge connection and an associated ghost field obeying the relations (\ref{gaugetransfo}). The obvious solution is to study a gauge theory of the AKSZ type, a condition fulfilled by the Chern-Simons model, the main subject of this paper. The BV formulation of the Wilson line contribution will happen to remain valid for other gauge theories.

So following \cite{AKSZObs}, the auxiliary fields are the maps between the odd tangent bundle of the curve $\Gamma$ and the coadjoint orbit,
\begin{displaymath}
\mathcal{F}^{\mathrm{aux}} = \mathrm{Map}(T\shone\Gamma,\mathcal{O}).
\end{displaymath}
This auxiliary space of fields needs to be equipped with its own BV structure, $\Omega^{\mathrm{aux}}$, that once added to the BV structure $\Omega^{\mathrm{amb}}$ of the ambient theory will provide the BV structure $\Omega = \Omega^{\mathrm{amb}} + \Omega^{\mathrm{aux}}$ of the full model with space of fields $\mathcal{F} = \mathcal{F}^{\mathrm{amb}} \oplus \mathcal{F}^{\mathrm{aux}}$. Then it will be possible to add to the ambient action $S^{\mathrm{amb}}$ an auxiliary term $S^{\mathrm{aux}}$ that obeys certain constraints to obtain a solution of the master equation of the full model.

The definition of $\Omega^{\mathrm{aux}}$ is similar to the one of the AKSZ-BV structure (\ref{AKSZBVstr}), we just need to change the source space and the symplectic structure of the target,
\begin{displaymath}
\Omega^{\mathrm{aux}} = \int_{T\shone \Gamma} \mu_\Gamma \, \tilde{\omega}_{\mathcal{O}}.
\end{displaymath}
Here $\mu_\Gamma$ is obviously the canonical measure on $T\shone\Gamma$.

Unfortunately, the Kirillov symplectic form on the coadjoint orbit is in general not exact, so it is in general not possible to find a Liouville one-form, which we would normally use to construct the kinetic term of the auxiliary action. However, in the case of integrable orbits, we may pick a line bundle (the pre-quantum line bundle in the language of geometric quantization) and a connection $\alpha_{\mathcal{O}}$ thereon with curvature $\omega_{\mathcal{O}}$,
\[
\delta\alpha_{\mathcal{O}} = \omega_{\mathcal{O}}
\]
(we recall that in the target spaces of AKSZ theories, we denote by $\delta$ the exterior derivative), and we can simply use this connection  to construct the kinetic term of the auxiliary action.

This formulation is not very practical to carry out calculations. To find expressions easier to deal with, we apply the defining property of the Kirillov symplectic form, that the pullback by the projection map
\begin{equation} \label{projectioncoadorbit}
\pi: G \rightarrow \mathcal{O}\simeq G/\mathrm{Stab}(T_0)
\end{equation}
brings it to an explicit presymplectic form $\omega_G$ on $G$,
\begin{equation} \label{defKirillov}
\pi^\ast (\omega_{\mathcal{O}}) = \omega_G = - \langle \mathrm{Ad}^\ast_g (T_0),\frac{1}{2}\left[\delta g\,g^{-1},\delta g\,g^{-1}\right]\rangle.
\end{equation}
This two-form is the contraction of $T_0$ with the exterior derivative of the Maurer-Cartan one-form on $G$. It thus admits a potential
\begin{equation}
\alpha_G = -\langle \mathrm{Ad}^\ast_g (T_0), \delta g\,g^{-1}\rangle.
\end{equation}
As it happens, the pullback by the projection map $\pi$ brings the connection $\alpha_{\mathcal{O}}$ over to the one-form $\alpha_G$,
\[
\pi^\ast (\alpha_\mathcal{O}) = \alpha_G,
\]
that we will use in the place of the more cumbersome connection to compute certain quantities. Since $\omega_G$ is degenerate, it is not possible to construct a Poisson bracket out of it, unless we restrict it to invariant functions on $G$, such as the ones obtained by pullback of functions on the coadjoint orbit $\mathcal{O}$ by the projection map $\pi$.

It remains to define the interaction term. The idea of \cite{AKSZObs} is to construct a function $\Theta_{\mathcal{O}}$ on $\mathfrak{g}\shone \times \mathcal{O}$ that will generate together with $\Theta_{\mathfrak{g}\shone}$ a Hamiltonian cohomological vector field on $\mathfrak{g}\shone \times \mathcal{O}$. While $\Theta_{\mathfrak{g}\shone}$ already satisfies an integrability condition on its own and generates a cohomological vector field $\mathcal{Q}_{\mathfrak{g}\shone}$, the integrability condition for  $\Theta_{\mathcal{O}}$ needs to be slightly adapted to account for the mixed term, namely
\begin{displaymath}
\mathcal{Q}_{\mathfrak{g}\shone} \Theta_{\mathcal{O}} + \frac{1}{2}\left\lbrace\Theta_{\mathcal{O}},\Theta_{\mathcal{O}}\right\rbrace_{\mathcal{O}} = 0.
\end{displaymath}
As it happens, the function
\begin{equation}
\Theta_{\mathcal{O}} = \langle \mathrm{Ad}^\ast_g (T_0), \psi\rangle
\end{equation}
satisfies this requirement and naturally extends the term $\langle\mathrm{Ad}^\ast_g (T_0), A \rangle$ that already appeared in the classical part (\ref{wilsonlineterm}). This integrability condition is most easily checked by pulling it back by $\pi$ to a function on $\mathfrak{g}\shone \times G$, where the Poisson bracket $\left\lbrace\cdot,\cdot\right\rbrace_{\mathcal{O}}$ becomes $\left\lbrace \cdot,\cdot\right\rbrace_G$, which can be explicitly determined by inverting $\omega_G$.

We now have all the ingredients to construct the auxiliary BV action, that we just need to combine into a formula similar as the usual AKSZ action (\ref{AKSZaction}),
\begin{displaymath}
S^{\mathrm{aux}} = \int_{T\left[1\right]\Gamma} \mu_\Gamma \ \left( \imath_{Q_D}\tilde{\alpha}_{\mathcal{O}} + \tilde{\Theta}_{\mathcal{O}} \right).
\end{displaymath}
By construction, if $S^{\mathrm{amb}}$ is the AKSZ action of the Chern-Simons model, the total action
\[
S=S^{\mathrm{amb}}+S^{\mathrm{aux}}
\]
automatically satisfies the classical master equation generated by the total BV structure
\[
\Omega = \Omega^{\mathrm{amb}} + \Omega^{\mathrm{aux}}.
\]
We claimed that it remains true when $S^{\mathrm{amb}}$ is the BV action of a generic gauge theory with gauge group $G$. To verify this assertion, we need to compute
\begin{equation} \label{threetermsCME}
\frac{1}{2}\left\lbrace S,S\right\rbrace =
\frac{1}{2}\left\lbrace S^{\mathrm{amb}},S^{\mathrm{amb}}\right\rbrace
+ \left\lbrace S^{\mathrm{amb}},S^{\mathrm{aux}}\right\rbrace
+ \frac{1}{2}\left\lbrace S^{\mathrm{aux}},S^{\mathrm{aux}}\right\rbrace.
\end{equation}
The first two terms involve only the ambient BV structure, since $S^{\mathrm{amb}}$ does not depend on the auxiliary fields. The first one vanishes due to the master equation of the BV ambient model. To compute the second term, we should know the exact dependence of the auxiliary term $S^{\mathrm{aux}}$ on the ambient fields $A$ and $\gamma$, and to compute the last one, we need an expression of the auxiliary BV structure $\Omega^{\mathrm{aux}}$ that we know how to invert.

These two issues can be addressed by using the projection map (\ref{projectioncoadorbit}) to define an extended space of fields,
\begin{displaymath}
\hat{\mathcal{F}}_G^{\mathrm{aux}} = \pi^\ast (\mathcal{F}^{\mathrm{aux}}) = \left\lbrace (g,g^+) \vert g\in \mathrm{Map}(\Gamma,G), g^+\in \Omega^1(\Gamma)\otimes g^\ast(T\mathcal{O})\left[-1\right] \right\rbrace.
\end{displaymath}
The subscript $G$ emphasizes the fact that the coadjoint orbit is replaced by the whole group.

This projection map, now seen as a map between spaces of fields,
\begin{displaymath}
\pi : \hat{\mathcal{F}}_G^{\mathrm{aux}} \rightarrow \mathcal{F}^{\mathrm{aux}},
\end{displaymath}
acts on the group-valued component $g$ by sending it to its image $\mathrm{Ad}^\ast_g(T_0)$ in the coadjoint orbit of $T_0$. It can be used to pull back differential forms on the auxiliary space of fields (such as the auxiliary BV structure, a two-form, or the auxiliary BV action, a zero-form) to this extended space of fields, where it is easier to compute BV brackets of $G$-invariant functionals given the explicit formulas for the pullbacks of the auxiliary BV structure and of the auxiliary action.

In $\hat{\mathcal{F}}_G^{\mathrm{aux}}$, both fields $g$ and $g^+$ can be combined into a superfield of total degree 0 that we can use to express the pullback of the auxiliary BV structure and action,
\begin{displaymath}
\mathbf{H}(x,\theta) = \mathrm{Ad}_{g(x)}^\ast (T_0) - \theta g^+(x).
\end{displaymath}
Here $x$ is a coordinate of $\Gamma$ and $\theta$ a Grassmanian coordinate on the odd fibers of $T\shone\Gamma$, and $g^+(x)$ is the component of the one-form $g^+$ expressed in this coordinate system, $g^+ = g^+(x)dx$.

We now have all the tools to compute the pullback of the auxiliary BV structure,
\begin{equation} \label{auxBVstr}
\begin{split}
\hat\Omega^{\mathrm{aux}}_G &= \pi^\ast(\Omega^{\mathrm{aux}}) = \int_{T\shone \Gamma} \mu_\Gamma \ \pi^\ast(\tilde{\omega}_{\mathcal{O}}) = \int_{T\shone \Gamma} \mu_\Gamma \ \tilde{\omega}_G \\
&= -\int_{T\shone \Gamma} \mu \, \delta\langle\mathbf{H},\delta g\,g^{-1}\rangle = \int_{\Gamma}\delta\langle g^+,\delta g\,g^{-1}\rangle \\
 &= \int_{\Gamma}\langle\delta g^+,\delta g\,g^{-1}\rangle + \langle g^+, \frac{1}{2}\left[\delta g\,g^{-1},\delta g\,g^{-1}\right]\rangle,
\end{split}
\end{equation}
and of the auxiliary BV action,
\begin{equation} \label{Saux}
\begin{split}
\hat S^{\mathrm{aux}}_G &= \pi^\ast(S^{\mathrm{aux}}) = \int_{T\left[1\right]\Gamma} \mu_\Gamma \ \left( \imath_{Q_D}\tilde{\alpha}_G + \pi^\ast(\tilde{\Theta}_{\mathcal{O}}) \right) \\
&= \int_{T\left[1\right]\Gamma} \mu_\Gamma \ \left( \langle \mathbf{H}, Dg\,g^{-1}
\rangle + \langle \mathbf{H},\mathbf{A}\rangle \right) \\
&= \int_{\Gamma} \left( \langle\mathrm{Ad}^\ast_g(T_0), A + dg\,g^{-1} \rangle - \langle g^+,\gamma\rangle\right).
\end{split}
\end{equation}
The additional term $\langle g^+,\gamma\rangle$ encodes the action of gauge transformations of the ambient model on the auxiliary classical field $\mathrm{Ad}^\ast_g(\langle T_0)$.

We insist on the fact that $\hat\Omega^{\mathrm{aux}}_G$ is only a pre-BV structure in $ \hat{\mathcal{F}}_G^{\mathrm{aux}}$, it is closed but degenerate. Nevertheless, like its finite dimensional counterpart $\omega_G$, it can be used to define a BV bracket on invariant functionals, such as the ones obtained by pullback from $\mathcal{F}^{\mathrm{aux}}$, for instance $\hat S^{\mathrm{aux}}_G$.

We can turn our attention to the pullback of the last two terms in (\ref{threetermsCME}). To compute the second one, we notice that $\left\lbrace S^{\mathrm{amb}},\cdot\right\rbrace$ acts as a differential (namely $Q$) on the fields of the ambient model, of which only $A$ and $\gamma$ appear in the auxiliary action (\ref{Saux}), so that we may simply use the relations (\ref{gaugetransfo}) to find
\begin{equation}
\begin{split}
\left\lbrace S^{\mathrm{amb}},\hat S^{\mathrm{aux}}_G\right\rbrace &= \int_{\Gamma} \left( \langle \mathrm{Ad}^\ast_g(T_0), Q(A) \rangle - \langle g^+,Q(\gamma)\rangle\right)  \\
&= \int_\Gamma \left( \langle \mathrm{Ad}^\ast_g(T_0), d_A\gamma \rangle -\langle g^+, \frac{1}{2}\left[ \gamma,\gamma\right] \rangle \right).
\end{split}
\end{equation}
Next, the bracket of the third term contains only contributions from the auxiliary structure, and we may compute
\begin{equation}
\begin{split}
\frac{1}{2}\left\lbrace \hat S^{\mathrm{aux}}_G,\hat S^{\mathrm{aux}}_G\right\rbrace &= \int_\Gamma \left( \langle g\frac{ \hat S^{\mathrm{aux}}_G \overleftarrow{\delta} }{\delta g}, \frac{\overrightarrow{\delta}\hat S^{\mathrm{aux}}_G}{\delta g^+}\rangle + \langle g^+, \frac{1}{2}\left[ \frac{ \hat S^{\mathrm{aux}}_G \overleftarrow{\delta} }{\delta g^+},\frac{\overrightarrow{\delta}\hat S^{\mathrm{aux}}_G}{\delta g^+}\right] \rangle \right) \\
&= \int_\Gamma \left( \langle (d+\mathrm{ad}_A^\ast)\mathrm{Ad}^\ast_{g^{-1}}T_0, \gamma\rangle + \langle g^+, \frac{1}{2}\left[ \gamma,\gamma\right] \rangle \right).
\end{split}
\end{equation}
The first line displays the BV bracket of invariant functionals on $\hat{\mathcal{F}}_G^{\mathrm{aux}}$ constructed out of the pre-BV structure $\hat{\Omega}^{\mathrm{aux}}_G$.

The sum of these two terms yields the integral of an exact term that vanishes since $\Gamma$ is closed (for now), and the pullback of the classical master equation is satisfied,
\begin{displaymath}
\pi^\ast \left( \frac{1}{2}\left\lbrace S^{\mathrm{amb}} + S^{\mathrm{aux}}, S^{\mathrm{amb}} + S^{\mathrm{aux}}\right\rbrace \right) = 0.
\end{displaymath}
Furthermore, since the left-hand side of this last equality is $G$-invariant, it behaves nicely enough under the projection $\pi$ so that $ S^{\mathrm{amb}} + S^{\mathrm{aux}}$ still solves the classical master equation at the level of $\mathcal{F} = \mathcal{F}^{\mathrm{amb}} \oplus \mathrm{Map}(T\shone\Gamma,\mathcal{O})$, also for generic gauge theories.
%We should add a short observation that if the BV space of fields admits a $\mathbb{Z}_2$-grading, which will happen when we discuss the one-dimensional Chern-Simons theory in section \ref{1DCSWL}, the BV bracket of the auxiliary action with itself will contain contributions from the ambient space of fields, since in that case $\gamma$ will also play the role of antifield for $A$, and some additionnal care will be needed.
\subsection{Quadratic Lie algebras}
In many cases of interest, the Lie algebra $\mathfrak{g}$ is equipped with a non-degenerate scalar product $(\cdot,\cdot)$, which we can use to define an isomorphism $\beta:\mathfrak{g}^\ast\rightarrow\mathfrak{g}$, that we can apply to $T_0$, $H$ and $g^+$. The relation $\beta(\mathrm{Ad}^\ast_g(T_0))=\Adg ( \beta(T_0))$ will be very useful, in particular we can replace the canonical pairing between $\mathfrak{g}$ and $\mathfrak{g}^\ast$ with the scalar product,
\begin{displaymath}
\langle \mathrm{Ad}^\ast_g(T_0), \cdot \rangle = \left(  \mathrm{Ad}_g(\beta(T_0)), \cdot \right),
\end{displaymath}
and thus identify coadjoint orbits with adjoint orbits.

In the rest of this article, we will assume that $\mathfrak{g}$ admits such a non-degenerate scalar product. We will make use of it to write down all actions and BV structures, and we will simply consider $T_0$, $H$ and $g^+$ to be elements of $\mathfrak{g}$ instead of its dual, $\mathfrak{g}^\ast$ (we drop the $\beta$ for simplicity). Coadjoint orbits will therefore be identified with adjoint orbits. To summarize the main results of this section with this new convention, we can re-write the auxiliary BV structure (\ref{auxBVstr}) as
\begin{equation}
\hat{\Omega}_G^{\mathrm{aux}} = \int_{\Gamma}(\delta g^+,\delta g\,g^{-1}) + ( g^+, \frac{1}{2}\left[\delta g\,g^{-1},\delta g\,g^{-1}\right])
\end{equation}
and the auxiliary BV action (\ref{Saux}) as
\begin{equation} \label{Sauxnew}
\hat{S}_G^{\mathrm{aux}} = \int_{\Gamma} \left( ( \mathrm{Ad}_g(T_0), A + dg\,g^{-1} ) - ( g^+,\gamma)\right).
\end{equation}
\section{3D Chern-Simons Theory with a Wilson line} \label{3DCSWL}
Our main goal in this paper is to study the behaviour of Chern-Simons models on a manifold $N$ with boundaries and Wilson lines ending on these boundaries, both in three and one dimension, in the BV formalism. The presence of a boundary requires either a careful choice in the boundary conditions for the fields, so as to keep the classical master equation under control, or the application of the recently developed BV-BFV formalism for gauge theories with boundaries \cite{CattaneoMnevReshetikhin}, which presents the big advantage that it allows to glue pieces together along their boundary.
%In short, one first finds the BV formulation of the bulk theory, namely a space of fields equipped with a BV structure $\Omega$ and a BV action $S$ satisfying the master equation $\left\lbrace S,S\right\rbrace =0$, which implies that the Hamiltonian vector field $Q$ it generates, $\imath_Q\Omega = \delta S$, is cohomological, $\left[Q,Q\right]=0$. In the presence of a boundary, this vector field is no longer Hamiltonian, due to surface terms related to the Liouville one-form $\alpha_\partial$ of the associated BFV structure of the boundary models. Essentially, this BFV structure is a symplectic structure $\Omega_\partial$ of degree 0 in the space of boundary fields (which simply are restrictions of the fields to their value on the boundary of the underlying manifold). Finally, the restriction $Q_\partial$ of the cohomological vector field $Q$ to the boundary will happen to be Hamiltonian with respect to the BFV structure, and the associated Hamiltonian functional will correspond to the BFV action $S_\partial$. Upon quantization, we will associate quantum states to the boundary of $N$ required to satisfy some gauge invariance condition, and the usual BV path integral in the bulk will lead to a transition amplitude.

In order to obtain the BV-BFV formulation of the three-dimensional Chern-Simons model with a boundary supporting some Wilson lines, we first need to determine the BV theory of the bulk. Actually we already know all its ingredients. Obviously, we will treat the Wilson lines as an auxiliary part of the action, as described in section \ref{WilsonLinesBV}, added to the ambient Chern-Simons action (\ref{Scsbv}),
\[
S^{\mathrm{amb}} = S_{BV}^{CS}.
\]
To include Wilson lines to our model, say $n$ of them, we need to extend the ambient space of fields
\begin{displaymath}
\mathcal{F}^{\mathrm{amb}} = \mathrm{Map}(T\shone N,\mathfrak{g}\shone)
\end{displaymath}
carrying the ambient BV structure $\Omega^{\mathrm{amb}} = \Omega^{CS}_{BV}$ with an auxiliary part
\begin{displaymath}
\mathcal{F}^{\mathrm{aux}} = \bigoplus_{k=1}^n \mathrm{Map}(T\shone\Gamma_k,\mathcal{O}_k)
\end{displaymath}
made of $n$ components, one for each Wilson line labeled by $k$. We recall that we denote by $\mathcal{O}_k$ the (co)adjoint orbit of a Lie algebra element $T_{0,k}$ encoding the representation in which the $k$-th Wilson line is computed and by $\Gamma_k$ the curve embedded in $N$ supporting this Wilson line. The BV structure of this auxiliary space of fields is the sum of $N$ copies of the auxiliary BV structure of a single Wilson line,
\begin{displaymath}
\Omega^{\mathrm{aux}} = \sum_{k=1}^n \int_{T\shone \Gamma_k} \mu_{\Gamma_k} \, \tilde{\omega}_{\mathcal{O}_k}.
\end{displaymath}
The auxiliary BV action is similarly constructed as a sum,
\begin{displaymath}
S^{\mathrm{aux}} = \sum_{k=1}^n \int_{T\shone \Gamma_k} \mu_{\Gamma_k} \ \left( \imath_{Q_D}\tilde{\alpha}_{\mathcal{O}_k} + \tilde{\Theta}_{\mathcal{O}_k} \right).
\end{displaymath}

From now on, unless specified otherwise, a superscript ``amb'' will always describe a quantity associated to the BV formulation of the bare Chern-Simons model, be it a BV structure, a BV action or a BV space of fields, ``aux'' will always describe a quantity associated to the BV formulation of the auxiliary contribution of $n$ Wilson lines, and no superscript will mean a BV quantity of the full model, namely $\mathcal{F}=\mathcal{F}^{\mathrm{amb}}\oplus\mathcal{F}^{\mathrm{aux}}$, $\Omega=\Omega^{\mathrm{amb}} + \Omega^{\mathrm{aux}}$ and $S=S^{\mathrm{amb}} + S^{\mathrm{aux}}$.

Before we consider the case of a source manifold with boundary, we need to compute the Hamiltonian vector field $Q$ generated by $S$, i.e. satisfying
\begin{equation} \label{iQOmega}
\imath_Q \Omega = \delta S.
\end{equation}
Once the BV structure $\Omega$ is inverted to form the BV bracket $\left\lbrace\cdot,\cdot\right\rbrace$, this is equivalent to
\[
Q = \left\lbrace S,\cdot\right\rbrace.
\]
This relation is linear in $Q$ and $S$, namely $\imath_{Q^{\mathrm{amb}}} \Omega = \delta S^{\mathrm{amb}}$ and $\imath_{Q^{\mathrm{aux}}} \Omega = \delta S^{\mathrm{aux}}$. We will again use the projection maps $\pi_k: G \rightarrow \mathcal{O}_k$ to pull back the auxiliary action and the auxiliary BV structure to the space of fields $\bigoplus_{k=1}^n \hat{\mathcal{F}}_G^{\mathrm{aux}}$ where the calculations are easier. Note that we need one copy of $\hat{\mathcal{F}}_G^{\mathrm{aux}}$ for each Wilson line. The results can then be easily brought over to the actual auxiliary space of fields by the $n$ projections $\pi_k$.

For the ambient Hamiltonian vector field we obtain
\begin{equation}
\begin{split}
Q^{\mathrm{amb}} = \left\lbrace S^{\mathrm{amb}}, \cdot \right\rbrace = & \left( \left( d_A A^+ + \left[ \gamma^+,\gamma\right]\right), \frac{\overrightarrow{\delta}}{\delta\gamma^+} \right) - \frac{1}{2} \left( \left[ \gamma, \gamma\right], \frac{\overrightarrow{\delta}}{\delta\gamma} \right) \\
& - \left( \left( dA + \frac{1}{2}\left[ A,A\right] + \left[ A^+,\gamma \right]\right), \frac{\overrightarrow{\delta}}{\delta A^+} \right) + \left( d_A \gamma, \frac{\overrightarrow{\delta}}{\delta A} \right),
\end{split}
\end{equation}
and for its auxiliary counterpart
\begin{equation}
\begin{split}
\hat{Q}_G^{\mathrm{aux}} = \left\lbrace \hat{S}_G^{\mathrm{aux}}, \cdot \right\rbrace = &  - \sum_k \left( \Adg(T_{0,k})\delta(\Gamma_k), \frac{\overrightarrow{\delta}}{\delta A^+} \right) - \left( d_A(\Adg(T_{0,k})), \frac{\overrightarrow{\delta}}{\delta g^+} \right) \\
& - \left( \gamma, g \frac{\overrightarrow{\delta}}{\delta g} \right)
- \sum_k \left( g^+ \delta(\Gamma_k),\frac{\overrightarrow{\delta}}{\delta \gamma^+} \right).
\end{split}
\end{equation}
Here $\delta(\Gamma_k)$ denotes a Dirac distribution two-form centred on $\Gamma_k$ to filter the curve out of the whole manifold $N$. The fields $g$ and $g^+$ that appear in front of these Dirac two-forms are defined only on the Wilson lines. The functional derivatives appearing right after them act on functionals in the bulk, but their results are zero- and one-forms that make sense on the curves $\Gamma_k$. Since the ambient  and the total actions solve classical master equations, we know that $Q^{\mathrm{amb}}$ and $Q=Q^{\mathrm{amb}}+Q^{\mathrm{aux}}$ are cohomological, but not $Q^{\mathrm{aux}}$.

If the source space has a boundary, on which might end an open Wilson line along one or more curves $\Gamma_k$ (for simplicity, we will assume all of them), these cohomological vector fields cease to be Hamiltonian due to boundary effects affecting the variation of the differentiated terms in the action. The integration by part required to compute the contribution of these terms to the variation of the action now contains a surface integral. The BV-BFV formalism is based on the observation that this correction can be seen as a one-form in the boundary space of fields $\mathcal{F}_\partial$. This boundary space of fields contains the restriction of the fields of the bulk BV theory to their value on the boundary of the source manifold, $\partial N$ in the case of our ambient theory, $\bigcup_{k=1}^n \partial \Gamma_k$ for the auxiliary model describing the Wilson lines. We denote by
\[
\pi_\partial: \mathcal{F} \rightarrow \mathcal{F}_\partial
\]
the projection corresponding to this restriction. The correction to the Hamiltonian condition (\ref{iQOmega}) can be expressed as
\begin{equation} \label{iQOmegaplusalpha}
\delta S = \imath_Q\Omega + \pi_\partial^\ast (\alpha_\partial).
\end{equation}
The exterior derivative of this one-form,
\[
\Omega_\partial = \delta\alpha_\partial,
\]
happens to be symplectic and is called the BFV structure. It is a two-form of ghost number 0 in the boundary space of fields. The corrected Hamiltonian condition (\ref{iQOmegaplusalpha}) is linear in $\alpha_\partial$, too, so we may decompose the boundary BFV structure
\[
\Omega_\partial = \Omega_\partial^{\mathrm{amb}} + \Omega_\partial^{\mathrm{aux}}
\]
and compute it in two parts. The ambient one corresponds to the Chern-Simons model,
\begin{equation} \label{2DBFVstrCS}
\Omega_\partial^{\mathrm{amb}} = \int_{\partial N} \left(  \frac{1}{2}\left( \delta A, \delta A \right) + \left( \delta \gamma, \delta A^+\right) \right),
\end{equation}
and could actually be derived in a two-dimensional adaptation of the AKSZ construction. To compute the auxiliary part, we make use of the usual trick to do calculations in the augmented space of fields. The result
\begin{equation}
\hat{\Omega}_{\partial,G}^{\mathrm{aux}} = \sum_k \int_{\partial\Gamma_k} \left( \mathrm{Ad}_g(T_{0,k}), \frac{1}{2}\left[ \delta g \, g^{-1}, \delta g \, g^{-1} \right] \right)
\end{equation}
is the sum of $2k$ copies of the symplectic form $\omega_G$ of the target space of the augmented space of fields $\hat{\mathcal{F}}^{\mathrm{aux}}_{\partial,G}$, one carried by each extremity of every Wilson line. Using the relation (\ref{defKirillov}), we immediately find the BFV structure on the actual auxiliary boundary space of fields $\mathcal{F}^{\mathrm{aux}}_\partial = \bigoplus_{k=1}^n \mathrm{Map}(\partial \Gamma_k, \mathcal{O}_K)$,
\begin{equation} \label{auxBFVstr}
\Omega_\partial^{\mathrm{aux}} = \sum_k \int_{\partial\Gamma_k} \tilde{\omega}_{\mathcal{O}_k},
\end{equation}
where $\tilde{\omega}_{\mathcal{O}_k}$ is evidently the Kirillov symplectic form on the $k$-th (co)adjoint orbit $\mathcal{O}_k$ lifted to the space of fields.

The curve $\Gamma_k$ being one dimensional, we have $\partial \Gamma_k = \left\lbrace z_k, z'_k\right\rbrace \subset \partial N$, and what the last integral really means is $\int_{\partial\Gamma_k} \tilde{\omega}_{\mathcal{O}_k} = \tilde{\omega}_{\mathcal{O}_k}(z_k) - \tilde{\omega}_{\mathcal{O}_k}(z'_k)$.

In the last step of the construction of the boundary BFV model, we know that the restriction of $Q$ to the boundary surface $\partial N$ is Hamiltonian with respect to the BFV structure, and the boundary BFV action is defined as its generating functional,
\begin{displaymath}
\imath_{Q_\partial}\Omega_\partial = \delta S_\partial.
\end{displaymath}
Ghost number counting shows that the BFV action has ghost number $1$.

In the case of the Chern-Simons model with Wilson lines, we calculate the restriction of $\hat{Q}_G = Q^{\mathrm{amb}} + \hat{Q}^{\mathrm{aux}}_G$ in the extended space of fields,
\begin{equation}
\begin{split}
\hat{Q}_{\partial,G}  = &  -\left( \frac{1}{2}\left[ \gamma, \gamma\right], \frac{\overrightarrow{\delta}}{\delta\gamma} \right)  - \left( \left( dA + \frac{1}{2}\left[ A,A\right] + \left[ A^+,\gamma \right]\right), \frac{\overrightarrow{\delta}}{\delta A^+} \right)\\
& + \left( d_A \gamma, \frac{\overrightarrow{\delta}}{\delta A} \right)  - \sum_k \left( \mathrm{Ad}_g(T_{0,k})\delta(\Gamma_k),\frac{\overrightarrow{\delta}}{\delta A^+} \right) - \left( \gamma, g \frac{\overrightarrow{\delta}}{\delta g} \right),
\end{split}
\end{equation}
which leads to the two contributions
\begin{equation}
S^{\mathrm{amb}}_{\partial} = - \int_{\partial N} \left(  \left( dA + \frac{1}{2}\left[ A,A\right] ,\gamma \right) + \left( A^+, \frac{1}{2}\left[ \gamma,\gamma\right] \right) \right)
\end{equation}
and
\begin{equation}
\hat{S}^{\mathrm{aux}}_{\partial,G} =  - \sum_k \int_{\partial \Gamma_k} \left( \mathrm{Ad}_g (T_{0,k}),\gamma\right)
\end{equation}
to the boundary BFV action. As expected, the $G$-valued field $g$ appears only in a (co)adjoint action, so the projection to the auxiliary space of fields is straightforward, and we obtain the BFV action
\begin{equation}
\begin{split} \label{BFVaction}
S_\partial =& - \int_{\partial N} \left(  \left( dA + \frac{1}{2}\left[ A,A\right] ,\gamma \right) + \left( A^+, \frac{1}{2}\left[ \gamma,\gamma\right] \right) \right. \\
& \quad +\left. \sum_k \left( \mathrm{Ad}_g (T_{0,k}),\gamma \right) (\delta^{(2)}(z_k) - \delta^{(2)}(z'_k)) \right).
\end{split}
\end{equation}
In the last line, we have cast everything into the integral over $\partial N$ by making use of Dirac distributions centered on the extremities of the Wilson lines.

We recognize in the boundary BFV action of the Chern-Simons model with Wilson lines an odd version of the two-dimensional $BF$ model with sources, where the role of the $B$ field is taken over by the restriction to the boundary of the ghost field $\gamma$ of the bulk theory.

We conclude this section with a short remark regarding the insertions (labeled by $z_k$ and $z'_k$) of the boundary model. In our setting, with Wilson lines ending on the boundary, these insertions always come in pairs of points carrying the same representation, one insertion at each end of a Wilson line. If we consider Wilson graphs in the bulk model, which are a natural generalizations of the Wilson lines, we can obtain any configuration of points and representations as insertions. Wilson graphs are observables modeled after Wilson lines, but based on oriented graphs instead of curves. Each edge carries a representation of $\mathfrak{g}$ and contributes with a similar term as a Wilson line to the total action, while each vertex carries an intertwining operator between the representations of the attached edges. If the formulation of these intertwining operators is straightforward in the operator formalism, their description is more involved in the path-integral formalism and goes beyond the scope of this paper, where we will for simplicity consider only Wilson loops and open Wilson lines.
\section{1D Chern-Simons Theory with a Wilson line} \label{1DCSWL}
The AKSZ construction for the Chern-Simons model can also be carried out in one dimension \cite{1DCSAlekseevMnev}. In this section, we will see how to add a Wilson line to this model, by following the same procedure as in the previous section. The main difference comes from the fact that the Wilson line is now a space-filling observable, and that the BV bracket of the auxiliary term with itself will pick up terms from the ambient part of the BV structure. Moreover, as stated before, we will now use a $\mathbb{Z}_2$-grading, since a $\mathbb{Z}$-grading is not possible in one dimension, so instead of denoting the ghost number in square brackets, we will use the parity-reversing operator $\Pi$.

Given a one-dimensional manifold $\Gamma$ (in general a disjoint union of circles and open segments), the space of fields is
\begin{equation}
\mathcal{F}^{\mathrm{amb}}=\mathrm{Map}( \Pi T\Gamma ,\Pi\mathfrak{g}),
\end{equation}
where $\mathfrak{g}$ is again assumed to be equipped with an invariant scalar product $\left(\cdot,\cdot\right)$.

The target space $\Pi\mathfrak{g}$ supports the same geometric structures as before, that we may again transpose to the space of fields.

If $x$ is a coordinate on $\Gamma$ and $\theta$ a Grassmanian coordinate on the odd fibers of $\Pi T\Gamma$, we can decompose the fields $\Psi\in\mathrm{Map}( \Pi T\Gamma ,\Pi\mathfrak{g})$ into a $\mathfrak{g}$-valued fermion $\psi$ and a $\mathfrak{g}$-valued one-form $A=A(x)dx$,
\begin{equation}
\Psi = \psi + \theta A(x).
\end{equation}
We repeat the same procedure to find the BV structure
\begin{equation} \label{1D_BV_form}
\Omega^{\mathrm{amb}}= - \int_{\Pi \Gamma}\mu \left( \delta\Psi,\delta\Psi \right) = \int_{\Gamma} \left( \delta\psi,\delta A \right)
\end{equation}
and the BV action
\begin{equation} \label{1D_action}
S^{\mathrm{amb}}  = \int_{\Pi \Gamma}\mu \left( \frac{1}{2}\left( \Psi, D\Psi\right) + \frac{1}{6}\left(\Psi,\left[ \Psi,\Psi\right]\right)\right) = \int_\Gamma \frac{1}{2} \left( \psi,d_A\psi\right).
\end{equation}
The $\mathfrak{g}$-valued one-form $A$ can be interpreted as a connection for some principal $G$-bundle over $\Gamma$, where $G$ is a Lie group integrating $\mathfrak{g}$. The odd $\mathfrak{g}$-valued scalar $\psi$ serves simultaneously as a ghost for the gauge symmetry and an antifield for $A$.

In order to add Wilson lines to this model, we need to extend the space of fields, the BV structure and the action with precisely the same auxiliary structure $\Omega^{\mathrm{aux}}$ and action $S^{\mathrm{aux}}$ as in the three-dimensional case, except that they are now supported directly by the base manifold of the ambient source space $\Gamma$. We consider a single Wilson line for simplicity, it is easy to add similar terms for additional lines. Furthermore we assume it covers the whole source space $\Gamma$. Actually it could involve only some of the connected components of $\Gamma$, and the other ones would support a bare (in the sense that there are no Wilson lines) one-dimensional Chern-Simons model. Notice also that instead of $\gamma$ we write $\psi$ to emphasize the fact that it plays simultaneously the role of $\gamma$ and $A^+$ of the previous model.

We insist once more that since the Wilson line is a space-filling observable, we need to check that the classical master equation is solved, a result which is not guaranteed by the AKSZ construction due to the term $\left\lbrace S^{\mathrm{aux}},S^{\mathrm{aux}}\right\rbrace_{\mathrm{ambient}}$ coming from the auxiliary part of the action and the ambient part of the BV structure. If we use again the projection map $\pi:G\rightarrow\mathcal{O}$ to pull differential forms from the auxiliary space of fields $\mathcal{F}^{\mathrm{aux}}$ to the extended one $\hat{\mathcal{F}}^{\mathrm{aux}}_G$, we can calculate
\begin{equation}
\begin{split}
\frac{1}{2}\left\lbrace \hat{S}_G,\hat{S}_G \right\rbrace &= \int_\Gamma \left(
\left( \frac{\hat{S}_G\overleftarrow{\delta}}{\delta A}, \frac{\overrightarrow{\delta}\hat{S}_G}{\delta \psi} \right)
+ \left( g\frac{\hat{S}_G\overleftarrow{\delta}}{\delta g}, \frac{\overrightarrow{\delta}\hat{S}_G}{\delta g^+} \right) \right.\\
& \qquad \quad \left. -\frac{1}{2} \left( \frac{\hat{S}_G\overleftarrow{\delta}}{\delta g^+},\left[g^+, \frac{\overrightarrow{\delta}\hat{S}_G}{\delta g^+}\right] \right) \right) \\
&= \int_\Gamma \left( -\frac{1}{2}\left[\psi,\psi\right] + \mathrm{Ad}_g T_0, d_A\psi + g^+\right) + \left( -d_A(\mathrm{Ad}_g T_0),-\psi \right) \\
& \qquad \quad - \frac{1}{2} \left( \psi, \left[ g^+,-\psi\right] \right) \\
&= \int_\Gamma \left( \mathrm{Ad}_g T_0,g^+ \right) \\
&= - \frac{1}{2}\int_{\Pi T\Gamma} \mu \left( \mathbf{H},\mathbf{H} \right) \\
&= - \frac{1}{2}\int_{\Pi T\Gamma} \mu \left( \frac{S^{\mathrm{aux}}\overleftarrow{\delta}}{\delta\Psi},\frac{\overrightarrow{\delta}S^{\mathrm{aux}}}{\delta\Psi} \right),
\end{split}
\end{equation}
and the last line shows it explicitly. Nevertheless, this term vanishes, since $g^+$ takes value in the tangent space $T_H\mathcal{O}$ at $H=\mathrm{Ad}_g T_0$ to the adjoint orbit, which is easily seen to be orthogonal to $H$ with respect to the invariant scalar product on $\mathfrak{g}$.

Again, before we turn to the case of a source space with a boundary, we need to compute the Hamiltonian cohomological vector field $Q$ generated by $S$, or more accurately its counterpart in the extended space of fields, namely
\begin{equation}
\begin{split}
\hat{Q}_G &= \left( \left( d_A\psi + g^+\right),\frac{\delta}{\delta A} \right) + \left( \left(-\frac{1}{2}\left[\psi,\psi\right] + \mathrm{Ad}_g(T_0)\right),\frac{\delta}{\delta\psi} \right) \\
& \quad - \left( \psi , g\frac{\delta}{\delta g} \right) - \left( \left(\left[\psi,g^+\right] + d_A(\mathrm{Ad}_g(T_0))\right),\frac{\delta}{\delta g^+}\right).
\end{split}
\end{equation}
If the source space has a boundary, in other words if some of its components are segments, we can repeat the procedure to construct the BFV boundary model. We first calculate the image of the symplectic potential of the boundary BFV structure in the augmented space of fields from the variation of the BV action,
\begin{equation}
\hat{\alpha}_{\partial,G} = \int_{\partial \Gamma} \left( \frac{1}{2} \left( \psi,\delta\psi \right) + \left( T_0, g^{-1}\delta g \right) \right),
\end{equation}
and the corresponding pre-BFV structure,
\begin{equation}
\hat{\Omega}_{\partial,G} = \int_{\partial \Gamma} \left( \frac{1}{2} \left( \delta\psi,\delta\psi \right)- \frac{1}{2} \left( \mathrm{Ad}_g(T_0), \left[\delta g\,g^{-1},\delta g\,g^{-1}\right] \right) \right) .
\end{equation}
We see that the second term is connected to the pullback by the projection map $\pi:G\rightarrow\mathcal{O}$ of the Kirillov-Kostant-Souriau symplectic structure, and we obtain as a BFV structure in the proper boundary space of fields
\begin{equation}
\Omega_\partial = \int_{\partial \Gamma} \left( \frac{1}{2} \left( \delta\psi,\delta\psi \right) + \tilde{\omega}_{\mathcal{O}} \right).
\end{equation}
In all these expressions, the integral over $\partial\Gamma$ is nothing but a sum over the boundary points, with each term carrying a sign given by the orientation of its segment.

The restriction of the cohomological vector field $\hat{Q}_G$ to the boundary,
\begin{equation}
\hat{Q}_{\partial,G} = \left( \left(-\frac{1}{2}\left[\psi,\psi\right] + \mathrm{Ad}_g (T_0)\right),\frac{\delta}{\delta\psi} \right) - \left( \psi , g\frac{\delta}{\delta g} \right) ,
\end{equation}
is Hamiltonian with respect to the BFV structure, and it is generated by the BFV action of the boundary model,
\begin{equation} \label{0dimBFV}
S_\partial = \int_{\partial I} \left( -\frac{1}{6} \left( \psi,\left[\psi,\psi\right] \right) + \left( \mathrm{Ad}_g (T_0),\psi \right) \right) .
\end{equation}
Finally we show that $S_\partial$ solves the master equation of the BFV model,
\begin{equation}
\left\lbrace S_\partial , S_\partial\right\rbrace = Q_\partial S_\partial = \int_{\partial\Gamma}\left( T_0,T_0 \right) = 0.
\end{equation}
As stated before, the last integral is really a sum over the boundary elements of $\partial\Gamma$ with a sign assigned to their orientation, and since they come in pairs, at each end of every segment, the overall sum vanishes.
\section{Boundary Quantum States} \label{BoundaryQuantumStates}
Upon quantization, the partition function and correlators of a field theory defined on a manifold $N$ without boundary are complex numbers. In the presence of a boundary, one should rather expect quantum states, elements of a Hilbert space associated to each component of the boundary $\partial N$, according to the Atiyah-Segal picture of quantum field theory. The disjoint union of boundary components corresponds to the tensor product of the associated Hilbert spaces. Then gluing together a pair of components of $\partial N$ corresponds to taking the scalar product of the two corresponding factors of the tensor product.

For instance, if $N=\left[ 0,1\right]$ is an interval, the partition function of the BV-BFV model should take value in some Hilbert space of the form $\mathcal{H}\otimes\mathcal{H}$, with one factor for each component of the boundary $\partial N =\left\lbrace 0,1\right\rbrace$, and upon gluing the two ends, contracting this tensor product using the scalar product on $\mathcal{H}$ should yield the BV partition function of the same model constructed on the circle $S^1$.

If the bulk theory is studied in the BV formalism, the boundary information is encoded in the associated BFV model, at least at the classical level, as we saw in the particular cases of the Chern-Simons theory in one and three dimensions, possibly with Wilson lines.

To pass to the quantum level, we first observe that a BFV boundary model can be canonically quantized. The BFV structure, a symplectic structure of ghost number 0 in the space of fields, is used to define the (anti)commutation rules for the quantized fields. These act on the Hilbert space $\mathcal{H}^{\mathrm{BFV}}_\partial$ associated to the boundary where the partition function of the bulk takes value. This Hilbert space inherits a grading from the ghost number of the classical fields. In this picture, the BFV action $S_\partial$, which was the generator of the cohomological vector field on the boundary $Q_\partial$, can be quantized by replacing the classical fields with their quantized counterparts, and we obtain the quantized BFV charge $\hat{S}_\partial$. Its action on the boundary space of states squares to zero and it roughly encodes the gauge transformations. At the classical level, a physical observable is a functional annihilated by the cohomological vector field $Q$ generated by the BV action in the bulk and the BFV action $S_\partial$ on the boundary, and two observables are gauge equivalent if they differ by a $Q$-exact term. At the quantum level, the role of $Q$ is taken over by the BFV charge $\hat{S}_\partial$: a gauge invariant boundary state should be annihilated by the BFV charge, and two states are gauge-equivalent if they differ by a BFV-exact term. Moreover, we require physical states to depend only on physical quantum fields, and not on the ghosts or the antifields. In other words, the space of boundary quantum states should correspond to the BFV-cohomology at ghost number zero $H^0_{\hat{S}_\partial}(\mathcal{H}^{\mathrm{BFV}}_\partial)$.

The relation with the quantized bulk theory is that the partition function (and all the other correlators) should obviously be gauge invariant and therefore belong to this cohomology $H^0_{\hat{S}_\partial}(\mathcal{H}^{\mathrm{BFV}}_\partial)$. Its determination thus becomes a subject of interest.

We will start with the one-dimensional Chern-Simons theory, a simpler model where all calculations can be done until the end, before we study the more interesting three-dimensional model.
\subsection{1D Chern-Simons Theory}
The zero-dimensional boundary model of the one-dimensional Chern-Simons theory contains $\mathfrak{g}$-valued fermions and bosonic fields $H=\mathrm{Ad}_g (T_0)$ which take value in the (co)adjoint orbit $\mathcal{O}$. Once quantized, the fermions form a Clifford algebra $Cl(\mathfrak{g})$. If $(t^a)_{a=1}^{\mathrm{dim}\mathfrak{g}}$ is an orthononormal basis of $\mathfrak{g}$ with structure constants $f_{abc}$, we obtain the anticommutation rules
\begin{equation}
\left[\hat\psi_a,\hat\psi_b\right]= \hbar\delta_{ab}
\end{equation}
for the quantized fermions.

For the bosonic content of the model, the Kirillov symplectic form on the (co)adjoint orbits is the inverse of the restriction from $\mathfrak{g}^\ast\simeq\mathfrak{g}$ to $\mathcal{O}$ of the Kirillov-Kostant-Souriau Poisson structure, so that the commutator of two $\mathcal{O}$-valued quantized fields is simply given by their Lie bracket. If we use the basis $(t_a)$ of the Lie algebra to write
\begin{displaymath}
H = \mathrm{Ad}_g(T_0) = X_a \, t_a,
\end{displaymath}
we can express the commutation rules with the structure constants of the Lie algebra,
\begin{equation} \label{bosonicquantumoperators}
\left[\hat X_a,\hat X_b\right]=\hbar f_{abc} \hat X_c.
\end{equation}
The corresponding sector of the algebra of quantum operators is a representation of the enveloping algebra $\mathcal{U}(\mathfrak{g})$ of the Lie algebra $\mathfrak{g}$, namely $\rho_R(\mathcal{U}(\mathfrak{g}))\subset \mathrm{End}(V_R)$. This representation is simply the representation $R$ in which we computed the Wilson loops in the previous sections.

We can use these operators $\hat\psi$ and $\hat X$ to construct the expectation value of the Wilson line $\langle W_{\Gamma, R}\rangle$ in the operator formalism, such as in \cite{1DCSAlekseevMnev}, where the partition function for the one-dimensional Chern-Simons model is derived in both the path-integral and the operator formalism. If the curve $\Gamma$ is open, this expectation value maps the space of fields to the boundary space of quantum states which is the cohomology at level 0 of the quantum BFV charge,
\begin{displaymath}
\langle W_{I, R}\rangle \in H^0_{\hat{S}_\partial}(\mathcal{H}_\partial).
\end{displaymath}
We need to find this cohomology.

The BFV charge
\begin{equation}
\hat{S}_\partial = \int_{\partial \Gamma} \hat{X}_a\hat{\psi}_a - \frac{1}{6} f_{abc}\hat{\psi}_a\hat{\psi}_b\hat{\psi}_c
\end{equation}
carries one copy of the cubic Dirac operator \cite{AlekseevMeinrenken}
\begin{displaymath}
\mathfrak{D} = \hat{X}_a\hat{\psi}_a - \frac{1}{6}f_{abc}\hat{\psi}_a\hat{\psi}_b\hat{\psi}_c
\end{displaymath}
at each boundary point of $\Gamma$. This operator squares to
\begin{displaymath}
\mathfrak{D}^2 = \frac{1}{2}\left[\mathfrak{D},\mathfrak{D}\right] = \frac{1}{2} \hat{X}_a\hat{X}_a - \frac{1}{48}f_{abc}f_{abc},
\end{displaymath}
a central element in the quantum Weil algebra $\mathcal{U}(\mathfrak{g})\otimes Cl(\mathfrak{g})$, which guarantees that the action of the BFV charge squares to zero.

It is known that this cohomology in trivial (\cite{AlekseevMeinrenken},\cite{1DCSAlekseevMnev}). The resulting quantized BV-BFV is therefore not very interesting, so we should turn to the more involved problem of the three-dimensional Chern-Simons theory.
\subsection{3D Chern-Simons Theory}
We may now repeat the same procedure for the three-dimensional model. The first observation is that the treatment of the part coming from the extremities of the Wilson lines, namely the terms in the insertion points labeled by $z_k$ and $z'_k$, is essentially the same as in the one-dimensional model. Each insertion contributes to the overall BFV structure with a term in (\ref{auxBFVstr}), that when canonically quantized gives the algebra of operators (\ref{bosonicquantumoperators}) we encountered in the quantization of the one-dimensional model. We can formally express the quantization map
\[
\mathrm{Ad}_{g(z_k)}(T_{k,0}) = X_{k,a}(z_k) t_a \mapsto \rho_k(\hat{X}_a(z_k)) t^a.
\]
Even though the orbits might be different for different insertions, the commutation rules (\ref{bosonicquantumoperators}) are identical for all of them, only the representation $\rho_k$ differs, as we emphasized on the right-hand side.

In the next step, if we choose a complex structure on the boundary surface $\Sigma=\partial N$, we get a polarization of the connection
\begin{equation}
A = A_z dz + A_{\overline{z}}d\overline{z},
\end{equation}
which allows us to rewrite the ambient part of the BFV structure (\ref{2DBFVstrCS}) in Darboux coordinates of the corresponding sector $\mathcal{F}^{\mathrm{amb}}_\partial$ of the BFV space of boundary fields,
\begin{equation}
\Omega^{\mathrm{amb}}_{\partial} = \int_{\partial N} dzd\overline{z} \left( \left( \delta A_z\delta A_{\overline{z}} \right) + \left( \delta\gamma,\delta A^+ \right) \right).
\end{equation}
Consequently, we may perform the canonical quantization by choosing among each pair of conjugated fields one quantum field and replace the other one by the corresponding functional differential, for instance
\begin{equation}
\begin{array}{rcl}
A_{\overline{z}} &\rightarrow & a, \\
A_z &\rightarrow & -\frac{\delta}{\delta a}, \\
\gamma &\rightarrow & \gamma, \\
A^+ &\rightarrow & \frac{\delta}{\delta\gamma},
\end{array}
\end{equation}
so as to obtain canonical (anti)commutation rules. Note that $a$ is a boson and $\gamma$ a fermion.

The Hilbert space $\mathcal{H}^{\mathrm{BFV}}_\partial$ of boundary states on which act all these operators is therefore the space of functionals in $a$ and $\gamma$ with value in a tensor product of all the representation space associated to each insertion,
\begin{displaymath}
\mathcal{H}^{\mathrm{BFV}}_\partial = \mathrm{Fun}\left(a,\gamma ; \bigotimes_k V_{\rho_k} \otimes V_{\rho_k}\right).
\end{displaymath}
We recall it is graded by the ghost number.

Among these states, we want to determine the cohomology $H^0_{\hat{S}_\partial}( \mathcal{H}^{\mathrm{BFV}}_\partial )$ of the BFV charge at ghost number zero, made up of the quantum states of the BV-BFV model.  At degree zero, we are considering functionals $\psi$ of the $\mathfrak{g}$-valued $(0,1)$-form $a$, independent of the ghosts $\gamma$, which take value in the tensor product of all representation spaces associated to the extremities of the Wilson lines of the models.

The BFV charge
\begin{equation}
\begin{split} \label{BFVcharge}
\hat S_\partial =& - \int_{\partial N}dzd\overline{z}
\left(
\left(\partial a + \overline{\partial}\frac{\delta}{\delta a} + \left[ a,\frac{\delta}{\delta a}\right],\gamma\right) \right. \\
& \quad - \left. \sum_k
\left(\rho_k(\hat{X}_a(z_k))\delta(z-z_k) - \rho_k(\hat{X}_a(z_{k'}))\delta(z-z_{k'})\right)
\left( t^a , \gamma\right)
\right. \\
& \quad + \left.
\left( \frac{1}{2}\left[ \gamma,\gamma\right], \frac{\delta}{\delta\gamma} \right)
\right)
\end{split}
\end{equation}
acts on the Hilbert space $\mathcal{H}^{\mathrm{BFV}}_\partial$ via multiplication and differentiation by the quantum fields $a$ and $\gamma$ and via the obvious action of the representation $\rho_k$ on its representation space $V_{\rho_k}$.

At ghost number zero, BFV quantum states $\psi$ are therefore subject to the condition
\begin{equation} \label{constraintquantumstates}
\left( \partial a + \overline{\partial}\frac{\delta}{\delta a} + \left[ a,\frac{\delta}{\delta a}\right] - \sum_k \left( \rho_k(\hat{X}_a(z_k))\delta_{z_k}(z) - \rho_k(\hat{X}_a(z_{k'}))\delta_{z_k}(z) \right)t^a \right) \psi = 0.
\end{equation}
This actually coincides with the constraint (1) in \cite{CSgenus0} imposed to the Schr\"{o}dinger picture states in the canonical quantization of the Chern-Simons model on $\Sigma\times\mathbb{R}$ at genus 0, or the constraint (2.2) in \cite{CSgenus1} in the same situation at genus 1, where it is found that the cohomology $H^0_{\hat{S}_\partial}(\mathcal{H}^{\mathrm{BFV}}_\partial)$ %corresponds to 
coincides with
the space of conformal blocks in the WZW model for a correlator of fields inserted at the extremities of the Wilson lines.

The condition (\ref{constraintquantumstates}) for quantum states also appears in the geometric quantization framework, see for instance constraint (3.4) in \cite{WittenCS}, therefore the space of states in geometric quantization coincides with the space of quantum boundary states in BV-BFV quantization.


\begin{thebibliography}{99}



%\bibitem{NonAbelianStokes} \dots

%\bibitem{CattaneoFelderPSM} A.~Cattaneo, G.~Felder, ``A path integral approach to the Kontsevich quantization formula '', Comm. Math. Phys. 212 (2000) 591-611
\bibitem{QuantSymplOrbits} A.~Alekseev, L.~Faddeev, S.~Shatashvili, ``Quantization of symplectic orbits of compact Lie groups by means of the functional integral'', J. Geom. and Phys.~5, 3 (1988) 391-406
\bibitem{AlekseevMeinrenken} A.~Alekseev, E.~Meinrenken, ``The non-commutative Weil algebra'', Invent. Math. 139 (2000) 135-172
\bibitem{1DCSAlekseevMnev} A.~Alekseev, P.~Mnev, ``One-dimensional Chern-Simons theory'', Comm. Math. Phys.~307 (2011) 185-227
\bibitem{AKSZ} M.~Alexandrov, M.~Kontsevich, A.~Schwarz, O.~Zaboronsky, ``The geometry of the master equation and topological quantum field theory'', Int.~J. Modern Phys.~A~12 (1997) 1405-1429
\bibitem{AKSZCattaneoFelder} A.~S.~Cattaneo, G.~Felder, ``On the AKSZ formulation of the Poisson sigma model'', Lett. Math. Phys. 56 (2001) 163-179
\bibitem{CattaneoMnevReshetikhin} A.~S.~Cattaneo, P.~Mnev, N.~Reshetikhin, ``Classical BV theories on manifolds with boundary'', 	arXiv:1201.0290
\bibitem{Diakonov-Petrov} D. Diakonov, V. Petrov, Sov. Phys. JETP Lett. 49 (1989) 284 ;
Phys. Lett. B 224 (1989) 131
\bibitem{CSgenus1} F.~Falceto, K.~Gawedzki,  ``Chern-Simons States at Genus One'', Comm. Math. Phys. 159 (1994) 549-580
\bibitem{Froehlich-King}  J. Fr\"ohlich, C. King, ``The Chern-Simons theory and knot polynomials'',
Comm. Math. Phys. 126, 1 (1989) 167-199
\bibitem{CSgenus0} K.~Gawedzki, A.~Kupiainen, ``$SU(2)$ Chern-Simons Theory at Genus Zero'', Comm. Math. Phys.~135 (1991) 531-546
\bibitem{OrbitMethod} A.~A.~Kirillov, ``Unitary representations of nilpotent Lie groups'', Doklady Akademii Nauk SSSR 138 (1961) 283-284
\bibitem{AKSZObs} P.~Mnev, ``A construction of observables for AKSZ sigma models'', in preparation
\bibitem{Roytenberg} D. Roytenberg, ``AKSZ-BV Formalism and Courant Algebroid-induced Topological Field Theories'', Lett. Math. Phys. 79 (2007) 143-159
\bibitem{Schwarz} A.~Schwarz, ``Geometry of Batalin-Vilkovisky quantization'', Comm. Math. Phys.~155 (1993) 249-260
\bibitem{WittenCS} E.~Witten, ``Quantum field theory and the Jones polynomial'', Comm. Math. Phys. 121, 3 (1989) 351-399
\end{thebibliography}
\end{document}